\documentclass[11pt]{article}
\usepackage{amssymb,amsmath,epsfig,verbatim}
\def\one{{{{\rm 1} \kern -.19em {\rm l}}}}
\def\C{{{{\rm {\mbox{\small l}}} \kern -.50em {\rm C}}}}
\def\R{{{{\rm l} \kern -.15em {\rm R}}}}
\def\N{{{{\rm l} \kern -.15em {\rm N}}}}
\def\E{{{{\rm l} \kern -.15em {\rm E}}}}
\def\P{{{{\rm l} \kern -.15em {\rm P}}}}
\def\Z{{{{\rm Z} \kern -.35em {\rm Z}}}}
\def\1{{{{\rm 1} \kern -.35em {\rm 1}}}}

\begin{document}
\begin{sloppypar}
\vspace*{0cm}
\begin{center}
{\setlength{\baselineskip}{1.0cm}{ {\Large{\bf
Dirac Equation In The Curved Spacetime and Generalized Uncertainty Principle: A fundamental quantum mechanical approach with energy dependent potentials
\\}} }}
\vspace*{1.0cm}
\"{O}zlem Ye\c{s}ilta\c{s} \\
Gazi University, Science Faculty, Department of Physics, 06500, Teknikokullar, Ankara, Turkey.
\end{center}
\noindent \\

\vspace*{.5cm}
\begin{abstract}
\noindent
In this work, we have obtained the solutions of the $(1+1)$ dimensional Dirac equation on a gravitational background within the  generalized uncertainty principle. We have shown that how minimal length parameters effect the Dirac particle in a   spacetime described by  conformally flat metric. Also, supersymmetric quantum mechanics is used both to factorize the Dirac Hamiltonians and  obtain new metric functions. Finally, it is observed that the energy dependent potentials may be extended to the energy dependent metric functions.
\end{abstract}

\section{Introduction}
Since 1938  much effort has been put into the construction of a well-established  theory of Quantum Gravity(QG). Some of the approaches to QG is string theory \cite{st}, loop quantum gravity \cite{lqg} and black hole \cite{bp} physics. According to the string theory, matter is composed of vibrating strings which have a fundamental size called as the Planck length $\ell_{Pl} \approx 10^{-35} m$  and at Planck scale,  a string cannot probe distances smaller than its length which evokes to a generalization of Heisenberg uncertainty principle to assocaite gravitational induced uncertainty. Within thought experiments and   derivations, it is suggested that the generalized uncertainty principle (GUP) holds at all scales. During the progress of this theory and different approaches, momentum space representation of quantum states has been considered by Kempf et al \cite{kempf}, the Dirac oscillator problem is analyzed in \cite{quesne}, the spectrum of the hydrogen atom in the presence of minimal length has been obtained in the coordinate space in \cite{S}. Modification of the Heisenberg algebra where the minimum observable length is zero results in the deformed algebra $[x,p]=i\hbar(1+\alpha x^{2}+\beta p^{2})$ which is also related to $\Delta x \Delta p \geq \frac{\hbar}{2}(1+\alpha(\Delta x)^{2}+\beta (\Delta p)^{2})$ \cite{kempf}, \cite{N}. When we consider the Planck energy scale, the Schr\"{o}dinger and Dirac equations of the whole quantum theory in GUP formalism are modified through the idea of a minimum observable length and a maximum observable momentum.  For the different methods of derivations of  modified uncertainty principle can be found in \cite{bp}. In \cite{N1}, it is shown that the  generalized form of momentum causes a generalized Dirac equation which shows that  in Planck scale, a free particle cannot be found because of the quantum oscillations.

The Dirac equation in low dimensions ($2+1$) has been intensely studied for the past three decades by many authors in theoretical physics \cite{moa}, \cite{sucu}, \cite{yes}, \cite{sha}.  The low dimensional systems have always been attracting attention due to theoretical and experimental interesting results such as high energy particle theory, condensed matter physics (monolayer structures), topological field theory, and string theory.  The algebra of Dirac spinors and matrices can be extended  to Riemannian spaces through a combination with the null tetrad formalism. Because a wide part of quantum mehanics in a curved spacetime is  the covariant Dirac equation formulation which   involves  arbitrary tetrad field on the spacetime. The tetrad field at any point  in spacetime, there is an orthonormal basis of the tangent space at that point. Recently, it has been investigated   that the Hamiltonian operator for  the covariant Dirac equation  depends on the choice of the tetrad field, hence the energy spectrum  also depend on that choice \cite{Ar}, \cite{ar}. Moreover, energy dependent potentials also arise in quantum mechanics within Levinson's theorem which gives an anology between the number of the bound states and a spherically symmetric short-range potential  producing energy dependent phase shift \cite{levi}, \cite{cal}. The energy dependent potentials find place in various branches of physics such as relativistic quantum mechanics \cite{ed1}, semiconductors \cite{ed2}, high energy physics \cite{ed3}. Among the investigation of mathematical complexities, one can be guided to the extension of the energy dependent potentials to the exceptional orthogonal polynomials  and $\mathcal{PT}$ symmetry \cite{sh1}, \cite{sh2}, and low dimensional Dirac Hamiltonians \cite{sh3}, \cite{sh4}.

In quantum mechanics, Hamiltonian hierarchies can be examined by Supersymmetric Quantum Mechanics (SUSYQM) proposing a symmetry exists between bosons and fermions \cite{junker}. These Hamiltonians come in pairs are known as partner Hamiltonians whose eigenstates sharing the same energy except the groundstate when the symmetry is unbroken. The algebraic tool, SUSYQM, is not limited to the study of non-relativistic quantum systems only but also applied to Pauli and Dirac Hamiltonians and their dynamics \cite{thaller}, \cite{quesne}, \cite{nieto}. Moreover, the geometrical exposition of quantum mechanics have been attracted much attention \cite{ar}.  The Dirac equation in the frame of a curved spacetime and the formulation was established by Fock and to Weyl, and it is referred to as the "Dirac-Fock-Weyl" (DFW) equation \cite{weyl}, \cite{fock}. This paper is devoted to the formulation of the Dirac equation in case of a minimal length which is also linked to a Dirac equation in a curved spacetime. The other strategy of the recent work is obtaining new Hamiltonian models via supersymmetric quantum mechanics which helps to generate the function of the conformally flat metric expression.

\section{Dirac equation and polynomial solutions}
This section shows that the Dirac equation in coordinate space within a  minimal length approach can be transformed into a a couple of differential equations via point canonical transformation. In the following Section, the Dirac equation in   one dimensional curved spacetime will be written in terms of tetrad fields using a conformally flat spacetime metric.

\subsection{generalized uncertainty principle and Dirac equation}
Different approaches to quantum gravity  give a prediction of a minimum measurable length, or a maximum observable momentum, and related modifications of the Heisenberg Uncertainty Principle. Now, the complete generalized uncertainty relation is defined to be \cite{kempf}
\begin{equation}\label{ze}
  \Delta x \Delta p \geq \frac{\hbar}{2}(1+\alpha (\Delta x)^{2}+\beta (\Delta p)^{2}+c)
\end{equation}
$\alpha, \beta$ and $c$ are positive constants. Then, the commutation of the position and momentum operators fulfill the relation given below
\begin{equation}\label{com}
  [x,p]=i\hbar (1+\alpha x^{2}+\beta p^{2}).
\end{equation}
There are also other forms of the GUP \cite{others}. For a special case of (\ref{ze}), we can consider
\begin{equation}\label{XP}
  [X,P]=i\hbar(1+\alpha X^{2})
\end{equation}
Then, the corresponding the  generalized uncertainty principle is \cite{kempf}
\begin{equation}\label{GUP}
  (\Delta X)(\Delta P) \geq \frac{\hbar}{2} (1+\alpha (\Delta X)^{2}).
\end{equation}
The operators $X$ and $P$  satisfying the canonical commutation relations can be given by \cite{meng}, \cite{messai},
 \begin{eqnarray}\label{XP1}
   X &=& x \\
   P &=& -i\hbar ((1+\alpha x^{2})\frac{d}{dx}+\gamma x),\label{XP2}
 \end{eqnarray}
where $\alpha$ is the minimal length parameter, $\gamma$ is an arbitrary constant. In that case, considering the one dimensional Dirac equation \cite{s}
\begin{equation}\label{eq}
 [ (E-V(X))-\sigma_{x}P_{x}-\sigma_{z}(m+S(X))]\phi(X)=0,
\end{equation}
$V(X)$ and $S(X)$ are the  scalar potentials. In terms of (\ref{XP1}) and (\ref{XP2}), and taking $\hbar=c=1$, (\ref{eq}) can be introduced as
\begin{eqnarray} \label{1}
  (1+\alpha x^{2})^{2}\phi''_{1}+\left(2x(1+\alpha x^{2})(\alpha+\gamma)-\frac{(1+\alpha x^{2})^{2}U'_{2}(x)}{E+U_{2}(x)}\right)\phi'_{1}+r_{1}(x)\phi_{1} &=& 0 \\
 (1+\alpha x^{2})^{2}\phi''_{2}+\left(2x(1+\alpha x^{2})(\alpha+\gamma)-\frac{(1+\alpha x^{2})^{2}U'_{1}(x)}{E+U_{1}(x)}\right)\phi'_{2}+r_{2}(x)\phi_{2} &=& 0 \label{2}
\end{eqnarray}
where
\begin{eqnarray}\label{Us}
  U_{1}(x) &=& -(V(x)+m+S(x)) \\
  U_{2}(x) &=& -(V(x)-m-S(x)), \label{Us1}
\end{eqnarray}
$\phi(x)=[\phi_{1}(x)~~~~i\phi_{2}(x)]^{T}$ ~~and we use
\begin{equation}\label{r1}
  r_{1}(x)=  E^{2}+\gamma(1+x^{2}(\alpha+\gamma))+E(U_{1}(x)+U_{2}(x))+EU_{1}(x)U_{2}(x)-\frac{x(1+\alpha x^{2})\gamma U'_{2}(x)}{E+U_{2}(x)},
\end{equation}
\begin{equation}\label{r2}
r_{2}(x)=E^{2}+\gamma(1+x^{2}(\alpha+\gamma))+E(U_{1}(x)+U_{2}(x))+EU_{1}(x)U_{2}(x)-\frac{x(1+\alpha x^{2})\gamma U'_{1}(x)}{E+U_{1}(x)}.
\end{equation}
By performing a point transformation as $r=\int \frac{dx'}{1+\alpha x'^{2}}$, we can express (\ref{1}) and (\ref{2}) as
\begin{equation}\label{3}
\begin{split}
  \phi''_{1}(r)+\left(2(\frac{\alpha+\gamma}{\sqrt{\alpha}}-\sqrt{\alpha})\tan\sqrt{\alpha} r-\frac{U'_{2}(r)}{E+U_{2}(r)}\right)\phi'_{1}(r)+(E^{2}+\gamma+\frac{\gamma(\alpha+\gamma)}{\alpha}(\tan\sqrt{\alpha }r)^{2}\\
  +E(U_{1}(r)+U_{2}(r)+U_{1}(r)U_{2}(r))-\frac{\gamma U'_{2}(r)\tan\sqrt{\alpha}r}{\sqrt{\alpha}(E+U_{2}(r))})\phi_{1}(r)=0
\end{split}
\end{equation}
\begin{equation}\label{4}
\begin{split}
  \phi''_{2}(r)+\left(2(\frac{\alpha+\gamma}{\sqrt{\alpha}}-\sqrt{\alpha})\tan\sqrt{\alpha} r-\frac{U'_{1}(r)}{E+U_{1}(r)}\right)\phi'_{2}(r)+(E^{2}+\gamma+\frac{\gamma(\alpha+\gamma)}{\alpha}(\tan\sqrt{\alpha }r)^{2}\\
  +E(U_{1}(r)+U_{2}(r)+U_{1}(r)U_{2}(r))-\frac{\gamma U'_{1}(r)\tan\sqrt{\alpha}r}{\sqrt{\alpha}(E+U_{1}(r))})\phi_{2}(r)=0.
\end{split}
\end{equation}
We also note that the inner product is defined to be as
\begin{equation}\label{ip}
  <\psi\mid\phi>=\int\frac{\psi^{\ast}\phi}{1+\alpha x^{2}}dx.
\end{equation}
%%%%%%%%%%%%%%%%%%%%%%%%%%%%%%%%%%%%%%%%%%%%%%%%%%%%%%%%%%%%%%%%%%%%%%%%%%%%%%%%%%%%%%%%%%%%%%%%%%%%%%%%%%%%%%%%%%%%%%%%%%%%%%%%%%%%%%%%%%%%%%%%%%%%%%%
%%%%%%%%%%%%%%%%%%%%%%%%%%%%%%%%%%%%%%%%%%%%%%%%%%%%%%%%%%%%%%%%%%%%%%%%%%%%%%%%%%%%%%%%%%%%%%%%%%%%%%%%%%%%%%%%%%%%%%%%%%%%%%%%%%%%%%%%%%%%%%%%%%%%%%%
\subsection{polynomial solutions}
We may search for the solutions of (\ref{3}) and (\ref{4}) here. After then, one can adapt (\ref{3}) to the Dirac system in a conformally flat spacetime later because of getting the results for this system. Now,  we give an ansatze for  $U_{2}(r)$ which reads
\begin{equation}\label{yutu}
  U_{2}(r)=-E+a \sin\sqrt{\alpha}r
\end{equation}
and for the sake of simplicity, the term inside of the parnethesis in the coefficient of $\phi_1(r)$ in (\ref{3}) can be chosen as a constant, i.e.
\begin{equation}\label{e}
 E(U_{1}(r)+U_{2}(r)+U_{1}(r)U_{2}(r))=c
\end{equation}
where $a,  c$ are constants. Hence, we can find $U_{1}(r)$ and other related energy dependent potentials $V(r)$ and $S(r)$ using (\ref{Us}), (\ref{Us1}), (\ref{yutu}) and (\ref{e}):
\begin{eqnarray}
  V(r) &=& \frac{c+E^{3}+aE\sin\sqrt{\alpha}r(-2E+a\sin\sqrt{\alpha}r)}{2E(-1+E-a\sin\sqrt{\alpha}r)}, \\
  S(r) &=& \frac{1}{2}(1-E-2m+a\sin\sqrt{\alpha}r+\frac{c+E}{E(-1+E-a\sin\sqrt{\alpha}r)}),
\end{eqnarray}
\begin{equation}\label{U1}
  U_{1}(r)=\frac{-c-E^{2}+aE\sin\sqrt{\alpha}r}{E(-1+E-a\sin\sqrt{\alpha}r)}.
\end{equation}
Then, performing a mapping given below
\begin{equation}\label{mp}
  \phi_{1}(r)=(\cos\sqrt{\alpha}r)^{\frac{\gamma}{\alpha}} \sqrt{\sin\sqrt{\alpha} r}\chi_{1}(r),
\end{equation}
leads to re-writing (\ref{3}) in the form of
\begin{equation}\label{c1}
  \chi''_{1}(r)+\left(c+E^{2}+\frac{\alpha}{4}-\frac{3\alpha}{4}\csc^{2}\sqrt{\alpha}r \right)\chi_{1}(r)=0.
\end{equation}
On the other hand, for (\ref{4}), one can use the transformation
\begin{equation}\label{chi}
  \phi_{2}(r)=\cos\sqrt{\alpha}r \sqrt{E+U_{1}(r)}\chi_{2}(r),
\end{equation}
and (\ref{4}) turns into
\begin{equation}\label{c2}
  \chi_{2}''(r)+\left(c+E^{2}-\frac{3(U'_{1}(r))^{2}}{4(E+U_{1}(r))^{2}}+\frac{U''_{1}(r)}{2(E+U_{1}(r))}\right)\chi_{2}(r)=0.
\end{equation}
If we turn back to our polynomial solutions, using  $\xi=\cos\sqrt{\alpha}r$, $\xi \in [0,1) $ in (\ref{c1})  gives
\begin{equation}\label{cc111}
  \xi(1-\xi)\chi_{1}''(r)+(\frac{1}{2}-\xi)\chi_{1}'(r)+(\varepsilon-\frac{3\alpha}{4}\frac{1}{1-\xi})\chi_{1}(r)=0,
\end{equation}
where $\varepsilon=\frac{1}{4}(c+E^{2}+\frac{\alpha}{4})$. Because there is singularity at $\xi=1$ in (\ref{cc111}), we seek for a solution of the type:
\begin{equation}\label{sol}
  \chi_{1}(\xi)= (1-\xi)^{\eta}y(\xi)
\end{equation}
and we get
\begin{equation}\label{sln}
  \xi(1-\xi)y''(\xi)+(\frac{1}{2}-\xi(1+2\eta))y'(\xi)+\frac{4\xi(\varepsilon-\eta^{2})+3\alpha-4\varepsilon+2\eta}{4(\xi-1)}y(\xi)=0.
\end{equation}
 It is noted that  Gauss hypergeometric equation is written as
\begin{equation}\label{Gauss}
  \xi(1-\xi)y''(\xi)+(\frac{1}{2}-(a+b+1)\xi)y'(\xi)-aby(\xi)=0.
\end{equation}
In order to get an equation like (\ref{Gauss}) from (\ref{sln}), we have to   get rid of the
last term in (\ref{sln}) which is the coefficient of $\frac{1}{\xi-1}$. This is satisfied when $\eta$ takes the special value which is given below
\begin{equation}\label{eta}
  \eta_{1,2}=\frac{1}{4}(1 \pm \sqrt{1+4c+\alpha}).
\end{equation}
Substituting (\ref{eta}) in (\ref{sln}) yields
\begin{equation}\label{final}
  \xi(1-\xi)y''(\xi)+(\frac{1}{2}-\xi(1+2\eta))y'(\xi)+(\varepsilon-\eta^{2})y(\xi)=0,
\end{equation}
we can also use the letters below:
\begin{eqnarray}
  a+b &=& 2\eta \\
 ab &=& \varepsilon-\eta^{2}\\
 \eta&=& \frac{1}{2}s\\
 s&=& \frac{1}{2}(1\pm \sqrt{1+4c+\alpha})~~~~s\geq 1.
\end{eqnarray}
Then, (\ref{Gauss}) has linear independent solutions \cite{mathbook}
\begin{eqnarray}
  y_{1}(\xi) &=& _{2}F_{1}(a,b,s+1/2,1-\xi)  \\
  y_{2}(\xi) &=&   (1-\xi)^{1/2-s}~~~ _{2}F_{1}(\frac{1}{2}-a, \frac{1}{2}-b, \frac{3}{2}-s, 1-\xi)
\end{eqnarray}
where $_{2}F_{1}(a,b,c,t)$ stands for the Gauss hypergeometric function. The parameters $a$ and $b$ are obtained as:
\begin{eqnarray}
  a &=&-\left( \frac{s}{2}-\frac{1}{2}\sqrt{c+E^{2}+\frac{\alpha}{4}}\right) \\
  b &=& \frac{s}{2}+\frac{1}{2}\sqrt{c+E^{2}+\frac{\alpha}{4}}.
\end{eqnarray}
Now, the solutions for $\phi_{1}(r)$ can be introduced as
\begin{equation}\label{phi1}
  \phi_{1,(1)}(r)= N_{(1)} (\cos\sqrt{\alpha})^{\frac{\gamma}{\alpha}}(\sin\sqrt{\alpha}r)^{1/2}(1-\cos\sqrt{\alpha}r)^{s/2}~~_{2}F_{1}(a,b,s+1/2,1-\cos\sqrt{\alpha}r)
\end{equation}
and
\begin{equation}\label{phi11}
  \phi_{1,(2)}(r)= N_{(2)} (\cos\sqrt{\alpha})^{\frac{\gamma}{\alpha}}(\sin\sqrt{\alpha}r)^{1/2(1-s)}(1-\cos\sqrt{\alpha}r)^{s/2}~~_{2}F_{1}(\frac{1}{2}-a, \frac{1}{2}-b, \frac{3}{2}-s, 1-\cos\sqrt{\alpha}r).
\end{equation}
If $\xi \in [0, 1)$, then, the boundary condition should be $\phi_{1}(x=\pm \pi)=0$, and a  physical wave function must vanish when $\xi\rightarrow 1$
\begin{equation}\label{bc}
 \lim _{\xi\rightarrow 1} ~~ \phi_{1}(\xi)=0.
\end{equation}
We remind that the Gauss hypergeometric function can be expanded about $u=0$ as \cite{mathbook}
\begin{equation}\label{gex}
  _{2}F_{1}(a,b,q,u)=1+\frac{ab}{q}\frac{u}{1!}+\frac{a(a+1)b(b+1)}{q(q+1)}\frac{u^{2}}{2!}+...
\end{equation}
then, the second linear independent solution $ \phi_{1,(2)}(r)$ is not bounded at $\xi=1$. So, we introduce the solution (\ref{phi1})
\begin{equation}\label{phi1r}
  \phi_{1}(r)= N_{1} (\cos\sqrt{\alpha}r)^{\frac{\gamma}{\alpha}}(\sin\sqrt{\alpha}r)^{1/2}(1-\cos\sqrt{\alpha}r)^{s/2}~~_{2}F_{1}(a,b,s+1/2,1-\cos\sqrt{\alpha}r),
\end{equation}
$N_1$ is a normalization constant. An analiticity condition for the hypergeometric function leads us to take
\begin{equation}\label{hyp}
  a=-j,~~j=0,1,2,...
\end{equation}
then we have,
\begin{equation}\label{phi1ra}
  \phi_{1,j}(r)= N_{1,j} (\cos\sqrt{\alpha}r)^{\frac{\gamma}{\alpha}}(\sin\sqrt{\alpha}r)^{1/2}(1-\cos\sqrt{\alpha}r)^{s/2}~~_{2}F_{1}(-j,b,s+1/2,1-\cos\sqrt{\alpha}r),
\end{equation}
and the quantization condition gives
\begin{equation}\label{en}
 -j=-(\frac{s}{2}-\frac{1}{2}\sqrt{c+E^{2}+\frac{\alpha}{4}})
\end{equation}
that turns out
\begin{equation}\label{energy}
  E_{j}=\pm \sqrt{(s-2j)^{2}-c-\frac{\alpha}{4}}.
\end{equation}
One can also calculate the solutions $\phi_{2}(r)$ using (\ref{eq}). These solutions can be written as
\begin{equation}\label{phi2}
\begin{split}
  \phi_{2}(r)=N_{2} \frac{(1-\cos\sqrt{\alpha}r)^{s/2}(\cos\sqrt{\alpha}r)^{-1+\gamma/\alpha}}{2a\sqrt{\alpha}(\sin\sqrt{\alpha}r)^{3/2}}
  (2\gamma ~_{2}F_{1}(a,b,1/2+s,1-\cos\sqrt{\alpha}r)(\sin\sqrt{\alpha}r)^{2}\\+\frac{(\cos\sqrt{\alpha}r)^{3}}{1+2s}((1+2s)~_{2}F_{1}(a,b,1/2+s,1-\cos\sqrt{\alpha}r))
  (s\alpha+(\alpha+s\alpha+2\gamma)\cos\sqrt{\alpha}r)-2\gamma \sec\sqrt{\alpha}r+\\  4ab\alpha~ _{2}F_{1}(1+a,1+b,3/2+s,1-\cos\sqrt{\alpha}r)\sin^{2}\sqrt{\alpha}r)(1+\alpha \tan^{2}\sqrt{\alpha}r).
  \end{split}
\end{equation}
\begin{figure}[h]
\begin{center}
\epsfig{file=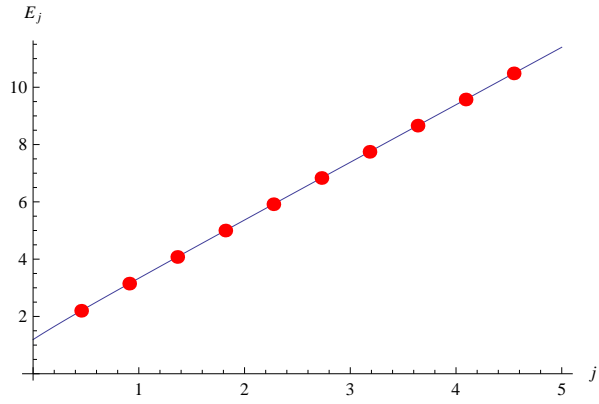,width=7.8cm}
\caption{The energy eigenvalues (\ref{energy}) for the values of  $\gamma=0.5$ and $\alpha=0.2$.}
\label{fig1}
\end{center}
\end{figure}
\begin{figure}[h]
\begin{center}
\epsfig{file= 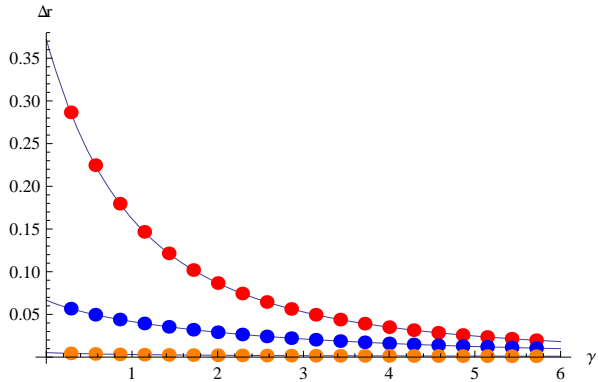,width=7.8cm}
\caption{The uncertainty in position versus $\gamma$ for the values of  $j=0$ (red), $j=1$ (blue) and $j=4$ (orange).}
\label{fig2}
\end{center}
\end{figure}
\\
It can be seen from Figure $1$ that for the positive values of the term inside square root which is  $(s-2j)^{2}-c-\alpha/4$, the energy increases with $j$. One can also calculate the uncertainty in position using the solutions $\phi_1(r)$ as given above in Figure $2$. For the growing quantum number $n$, the uncertainty in position decreases, especially for the greatest values of  $\gamma$, it is decaying to zero.
%%%%%%%%%%%%%%%%%%%%%%%%%%%%%%%%%%%%%%%%%%%%%%%%%%%%%%%%%%%%%%%%%%%%%%%%%%%%%%%%%%%%%%%%%%%%%%%%%%%%%%%%%%%%%%%%%%%%%%%%%%%%%%%%%%%%%%%%%%%%%%%%%%%%%%%%%%%%%%%%%%
%%%%%%%%%%%%%%%%%%%%%%%%%%%%%%%%%%%%%%%%%%%%%%%%%%%%%%%%%%%%%%%%%%%%%%%%%%%%%%%%%%%%%%%%%%%%%%%%%%%%%%%%%%%%%%%%%%%%%%%%%%%%%%%%%%%%%%%%%%%%%%%%%%%%%%%%%%%%%%%%%
\section{Dirac equation in $(1+1)$ dimensional curved spacetime}
When the Weyl tensor for the spacetime metric vanishes, then, metric allows conformal flatness. A non-static conformally flat metric can be written as
\begin{equation}\label{ds}
  ds^{2}=\Omega(x,t)^{2}(dt^{2}-dx^{^{2}})
\end{equation}
where $\Omega(x,t)$ is a function of position and time. The Dirac equation  in the background of a curved
spacetime is:
\begin{equation}\label{D1}
[ i \gamma^{\mu}(x)(\nabla_{\mu}+i e A_{\mu})-(m+\Delta(x))]\Psi(\vec{x})=0,
\end{equation}
where the local $\gamma$ matrices are being transformed as $\gamma^{\mu}=e_{a}^{\mu}(x) \gamma^{a}$, $[\gamma^{\mu}(x), \gamma^{\nu}(x)]=2g^{\mu \nu}(x)$, $A_{\mu}(x)$ is the vector potential and $\Delta(x)$ is a scalar potential.  It is convenient to introduce a generally covariant derivative $\nabla_{\mu}$ as
\begin{equation}\label{cd}
  \nabla_{\mu}=\frac{\partial}{\partial x^{\mu}}+\Gamma_{\mu}
\end{equation}
where the spin connection is
\begin{equation}\label{Gam}
  \Gamma_{\mu}=-\frac{1}{8}[\gamma^{\mu},\gamma^{\nu}] g_{ab} e^{a}_{\mu}\nabla_{\mu} e^{b}_{\nu}.
\end{equation}
Here, $e_{a}^{\mu}(x)$ are the vielbeins which are $e_{0}^{0}=e_{1}^{1}=\frac{1}{\Omega}$. The spin connection expression can be obtained as requiring that both objects have the same covariant derivative, we get the constraint  $\nabla_{\nu} e^{\mu}_{a}=0$,
\begin{equation}\label{ome}
  \omega_{b \nu}^{a}=e_{\mu}^{a} \partial_{\nu}(e_{b}^{\mu})+e_{\mu}^{a} e_{\sigma}^{b}\Gamma_{\sigma \nu}^{\mu}.
\end{equation}
We can also remind the spinor representation
\begin{equation}\label{sr}
  \nabla_{\mu}\Psi=(\partial_{\mu}+\frac{1}{4}\omega^{ab}_{\mu}(x)\gamma_{ab})\Psi.
\end{equation}
Then, the spin connections are $\omega_{10}^{0}=\omega_{00}^{1}=\frac{\Omega_{x}}{\Omega}$ and $\omega^{0}_{11}=\omega_{01}^{1}=\frac{\dot{\Omega}}{\Omega}$. Here we use the partial derivatives as $\Omega'=\Omega_{x}=\frac{\partial \Omega}{\partial x}$ and $\dot{\Omega}=\frac{\partial \Omega}{\partial t}$. We also note that the Christoffel symbols which are different from zero are: $\Gamma_{00}^{0}=\Gamma^{0}_{11}=\Gamma^{1}_{10}=\Gamma^{1}_{01}=\frac{\dot{\Omega}}{\Omega}$ and $\Gamma^{0}_{01}=\Gamma^{0}_{10}=\Gamma^{1}_{00}=\Gamma^{1}_{11}=\frac{\Omega'}{\Omega}$. Now, (\ref{D1}) can be multiplied by $\Omega \gamma^{0}$ from the left, we get
\begin{equation}\label{D3}
  i\left(\partial_{t}+\gamma^{0}\gamma^{1}\frac{[\gamma^{0},\gamma^{1}]}{4}\frac{\dot{\Omega}}{\Omega}\right)\Psi=i\left(\gamma^{0}\gamma^{1}\partial_{x}+ie\gamma^{0}\gamma^{1} A_{1}(x)+\frac{[\gamma^{0},\gamma^{1}]}{4}\frac{\Omega_{x}}{\Omega}\right)\Psi+\gamma^{0}\Omega (m+\Delta(x))\Psi.
\end{equation}
Using $\gamma^{0}=\sigma_{z}$, $\gamma^{1}=-i\sigma_{y}$, one obtains
\begin{equation}\label{D4}
  i \partial_{t}\Psi=\left(\sigma_{x}(P_{x}+W(x))-\sigma_{z}\Omega(m+\Delta(x))\right)\Psi
\end{equation}
where we take $\dot{\Omega}=0$ and use
\begin{equation}\label{w}
  W(x)= e A_{1}(x)-i\frac{\Omega_{x}}{2\Omega}.
\end{equation}
Introducing $\Psi(x,t)$ as
\begin{equation}\label{psi}
  \Psi(x,t)=e^{-iEt}~ e^{\int^{x} \rho(z)dz}~ [\psi_{1}(x)     ~~~~ i\psi_{2}(x)]^{T}
\end{equation}
then, (\ref{D4}) gives the couple of first order differential equations
\begin{eqnarray}\label{D5}
  \left[E+\Omega(x)(m+\Delta(x))\right]\psi_{1}(x)&=& \left(\frac{d}{dx}+iW(x)+\rho(x)\right)\psi_{2}(x) \\
 \left[E-\Omega(x)(m+\Delta(x))\right]\psi_{2}(x) &=&\left(-\frac{d}{dx}-iW(x)-\rho(x)\right)\psi_{1}(x). \label{D500}
\end{eqnarray}
Here we use $\Phi(x)=\Omega(x)(m+\Delta(x))$. Hence, we may get
\begin{equation}\label{D50}
  -\psi''_{1}-F_{-}(x)\psi'_{1}(x)+(W(x)^{2}-iW'(x)+\Phi(x)^{2}-2iW(x)\rho(x)-\rho(x)^{2}-\rho'(x)-\frac{(iW(x)+\rho(x))\Phi'(x)}{E-\Phi(x)}-E^{2})\psi_{1}(x)  =  0
\end{equation}
\begin{equation}\label{D51}
  -\psi''_{2}-F_{+}(x)\psi'_{2}(x)+(W(x)^{2}-iW'(x)+\Phi(x)^{2}-2iW(x)\rho(x)-\rho(x)^{2}-\rho'(x)+\frac{(iW(x)+\rho(x))\Phi'(x)}{E+\Phi(x)}-E^{2})\psi_{2}(x) =  0
\end{equation}
where $F_{\mp}(x)=(-2i(W(x)-i\rho(x))\mp \frac{\Phi'(x)}{E\mp\Phi(x)})$. It is noted that (\ref{phi1ra}) can also be the solutions of Equation (\ref{D50}); $\phi_{1}(r)=\psi_{1}(x\rightarrow r)$ under some restrictions for (\ref{D50}).  So, in (\ref{D50}), applying $x\rightarrow r$ and using (\ref{3}), (\ref{c1}) and (\ref{D50}), we have
\begin{equation}\label{pWp}
  -2iW(r)-2\rho(r)- \frac{\Phi'(r)}{E-\Phi(r)}-\sqrt{\alpha}\cot\sqrt{\alpha}r+\frac{2\gamma}{\sqrt{\alpha}}\tan\sqrt{\alpha}r=0
\end{equation}
\begin{equation}\label{pWp1}
  c-E^{2} - iW'(r)+W(r)^{2}+\Phi(r)^{2}+(\gamma+\frac{\gamma^{2}}{\alpha})\tan^{2}\sqrt{\alpha}r-\rho^{2}(r)-\rho'(r)-\frac{(\rho(r)+iW(r))\Phi'(r)}{E-\Phi(r)}=0.
\end{equation}
From (\ref{pWp}) and (\ref{pWp1}), we get the solutions $W(r)$, $\rho(r)$ and $\Phi(r)$
\begin{eqnarray}\label{D6}
  \Phi(r) &=& E(1+C_1) \\
  W(r) &=& -\frac{i\gamma}{\sqrt{\alpha}}\tan\sqrt{\alpha}r \\ \label{D7}
  \rho(r)&=& -\frac{\sqrt{\alpha}}{2} \cot\sqrt{\alpha}r \label{D8}
\end{eqnarray}
where $C_1$ is a constant and it can be taken as $E(1+C_1)=1$. Now we can note that (\ref{energy})  is also the spectrum of the system given in (\ref{D50}) as long as (\ref{pWp}), (\ref{pWp1}) are satisfied. Then,  $\Omega(x)=1$ gives us a flat spacetime. Here, it can be noted that $\phi_{2}(r)\neq \psi_{2}(x\rightarrow r)$. Then, $\psi_{2}(r)$ can be obtained using
\begin{equation}\label{de1}
  \psi_2(r)= \frac{1}{E-\Phi(r)}\left(-\frac{d}{dr}-iW(r)\right)\psi_1(x).
\end{equation}
We remind that the solutions $\psi_{1}(r)$ are given by
\begin{equation}\label{psi1}
  \psi_{1}(r)=N_{1} (\cos\sqrt{\alpha}r)^{\frac{\gamma}{\alpha}}\sqrt{\sin\sqrt{\alpha}r}(1-\cos\sqrt{\alpha}r)^{s/2} ~_{2}F_{1}(-j, b, s+1/2, 1-\cos\sqrt{\alpha}r)
\end{equation}
and $\psi_{2}(r)$ are found as
\begin{equation}\label{psi2}
\begin{split}
   \psi_2(r)= \frac{N_2}{E-\Phi(r)}\left(\frac{(1-\cos\sqrt{\alpha}r)^{s/2}(\cos\sqrt{\alpha}r)^{\gamma/\alpha}}{2(1+2s)\sqrt{\alpha}\sqrt{\sin\sqrt{\alpha}r}}\right)\\
   ((-1+2s)~_{2}F_{1}(b, -j,1/2+s, 1-\cos\sqrt{\alpha}r)(s\alpha+(c+\alpha+s\alpha+2\gamma)\cos\sqrt{\alpha}r-(c+2\gamma)\sec\sqrt{\alpha}r)&+\\
   4bj\alpha~_{2}F_{1}(1+b,1-j,3/2+s,1-\cos\sqrt{\alpha}r)\sin^{2}\sqrt{\alpha}r).
   \end{split}
\end{equation}
We note that the solutions (\ref{psi1}) and (\ref{psi2}) are the components of (\ref{psi}) when $x \rightarrow r$.
%%%%%%%%%%%%%%%%%%%%%%%%%%%%%%%%%%%%%%%%%%%%%%%%%%%%%%%%%%%%%%%%%%%%%%%%%%%%%%%%%%%%%%%%%%%%%%%%%%%%%%%%%%%%%%%%%%%%%%%%%%%%%%%%%%%%%%%%%%%%%%%%%%%%%%%%%%%%
%%%%%%%%%%%%%%%%%%%%%%%%%%%%%%%%%%%%%%%%%%%%%%%%%%%%%%%%%%%%%%%%%%%%%%%%%%%%%%%%%%%%%%%%%%%%%%%%%%%%%%%%%%%%%%%%%%%%%%%%%%%%%%%%%%%%%%%%%%%%%%%%%%%%%%%%%%
\subsection{supersymmetric analysis}
We can search for partner Hamiltonians of our system. Choosing the superpotential $\textbf{W}(r)=C_1 \cot\sqrt{\alpha}r$, we can handle (\ref{c1}) as given below
\begin{equation}\label{h1}
  h_{1}=A^{\dag}A,~~~~A=\frac{d}{dr}+\textbf{W}(r),
\end{equation}
\begin{equation}\label{h1h}
  h_{1} \chi_{1}=-\frac{d^{2}\chi_1}{dr^{2}} +(-c-\frac{\alpha}{4}+\frac{3\alpha}{4}\csc^{2}\sqrt{\alpha}r)\chi_1=E^{2}\chi_{1}
\end{equation}
where $h_1$ is the corresponding Hamiltonian and $C_1$ is a parameter. If we compare (\ref{h1h}) with (\ref{c1}), we find
 \begin{eqnarray}
   c  &=& 2 \alpha  \\
   C_1 &=&- \frac{3\sqrt{\alpha}}{2}.
 \end{eqnarray}
The Hamiltonian $H_1$ corresponding to the system  in  (\ref{3}) reads,
\begin{equation}\label{H1}
  H_1=-\frac{d^{2}}{dr^{2}}+(\sqrt{\alpha}\cot\sqrt{\alpha}r-\frac{2\gamma}{\sqrt{\alpha}}\tan\sqrt{\alpha}r)\frac{d}{dr}+
  (-2\alpha-\frac{\gamma(\alpha+\gamma)}{\alpha}\tan^{2}\sqrt{\alpha}r),
\end{equation}
then we can continue with noting that $h_1$ and $H_1$ are isospectral, i.e.
\begin{equation}\label{chain}
 ( \phi_{1}\rho)^{-1} h_1 \rho \phi_1=E^{2},~~~~\rho(r)=(\cos\sqrt{\alpha}r)^{-\frac{\gamma}{\alpha}}\sqrt{\csc\sqrt{\alpha}r}.
\end{equation}
The solutions of the system in (\ref{h1h}) is already known which is (\ref{energy}). Now we can generate the partner of the Hamiltonian $H_1$ which is $H_2$ using $h_{2}=AA^{\dag}$,
\begin{equation}\label{h2}
  h_{2}=-\frac{d^{2}}{dr^{2}}-\frac{9\alpha}{4}+\frac{15\alpha}{4}\csc^{2}\sqrt{\alpha}r,~~~~h_{2}\chi^{(2)}_{1}=E^{2}\chi^{(2)}_{1}
\end{equation}
where $\chi^{(2)}_{1}$ is the wavefunction of the Hamiltonian $h_{2}$  which is not equal to $\chi_{2}(r)$. Then, using (\ref{h2}) we can find $H_{2}$ Hamiltonian which is the partner Hamiltonian of (\ref{H1}) as:
\begin{equation}\label{H2}
  H_2=-\frac{d^{2}}{dr^{2}}+(\sqrt{\alpha}\cot\sqrt{\alpha}r-\frac{2\gamma}{\sqrt{\alpha}}\tan\sqrt{\alpha}r)\frac{d}{dr}+
  (-2\alpha+\gamma+3\alpha \csc^{2}\sqrt{\alpha}r-\gamma \sec^{2}\sqrt{\alpha}r).
\end{equation}
If we think back on previous topic, one can wonder that if another $\Omega(r)$ which is non-constant could be defined for the system (\ref{H2}). If $H_{1}$ is related to $\Omega(r)=1$, then, can we find a different $\Omega(r)$ for $H_{2}$?  Let us look at $H_{2}$ in (\ref{H2}) and (\ref{D51}). We have following equations:
\begin{equation}\label{pWpWp}
  -2iW(r)-2\rho(r)+ \frac{\Phi'(r)}{E+\Phi(r)}-\sqrt{\alpha}\cot\sqrt{\alpha}r+\frac{2\gamma}{\sqrt{\alpha}}\tan\sqrt{\alpha}r=0
\end{equation}
\begin{equation}\label{pWp1W}
   - iW'(r)+W(r)^{2}+\Phi(r)^{2}+-2\alpha+\gamma+3\alpha \csc^{2}\sqrt{\alpha}r-\gamma \sec^{2}\sqrt{\alpha}r-\rho^{2}(r)-\rho'(r)+\frac{(\rho(r)+iW(r))\Phi'(r)}{E+\Phi(r)}=0.
\end{equation}
Following the same procedure, it can be seen that  $\Omega(r)$ solutions are also a constant for $H_2$. We also remind that the relation between the energy levels and eigenstates   of the systems $h_1$ and $h_2$ as given below
\begin{equation}\label{E12}
  \epsilon^{(1)}_{j+1}=\epsilon^{(2)}_{j},~~~~\epsilon^{(1)}_{0}=0
\end{equation}
where $E^{2}=\epsilon$, $\epsilon^{(1)}_{j}$ and $\epsilon^{(2)}_{j}$ are corresponding to the energy eigenvalues of the systems $h_1$ and $h_2$ respectively. \begin{equation}\label{wf12}
  \chi^{(1)}_{j+1}=(\epsilon^{(2)}_{j})^{-1/2}A^{\dag}\chi^{(2)}_{j},~~~~\chi^{(1)}_{j+1}=\chi_{1,j+1}.
\end{equation}
Henceforth, we will continue with a point transformation for the system (\ref{D5})-(\ref{D500}) because our task is obtaining a non-constant $\Omega(r)$. We use
\begin{equation}\label{new}
   \frac{dx}{dy}=F(y)
\end{equation}
and (\ref{new}) leads to
\begin{eqnarray}\label{D55}
  \left[E+\Omega(y)(m+\Delta(y))\right]\psi_{1}(y)&=& \left(\frac{1}{F(y)}\frac{d}{dy}+iW(y)+\rho(y)\right)\psi_{2}(y) \\
 \left[E-\Omega(y)(m+\Delta(y))\right]\psi_{2}(y) &=&\left(-\frac{1}{F(y)}\frac{d}{dy}-iW(y)-\rho(y)\right)\psi_{1}(y). \label{D5500}
\end{eqnarray}
Then, (\ref{D55}) and (\ref{D5500}) can be evolved as
\begin{equation}\label{eqq1}
  \begin{split}
  -\psi''_{1}(y)+(-2F(y)(\rho(y)+iW(y))+\frac{F'(y)}{F(y)}-\frac{\Phi'(y)}{E-\Phi(y)})\psi'_{1}(y)+\\
  (F(y)^{2}(-E^{2}+W(y)^{2}-2iW(y)\rho(y)-\rho(y)^{2}+\Phi(y)^{2})-F(y)(iW'(y)+\rho'(y)+\\
  \frac{iW(y)\Phi'(y)}{E-\Phi(y)}+\frac{\rho \Phi'(y)}{E-\Phi(y)}))\psi_{1}(y)=0
  \end{split}
\end{equation}
\begin{equation}\label{eqq2}
  \begin{split}
  -\psi''_{2}(y)+(-2F(y)(\rho(y)+iW(y))+\frac{F'(y)}{F(y)}+\frac{\Phi'(y)}{E+\Phi(y)})\psi'_{2}(y)+\\
  (F(y)^{2}(-E^{2}+W(y)^{2}-2iW(y)\rho(y)-\rho(y)^{2}+\Phi(y)^{2})-F(y)(iW'(y)+\rho'(y)-\\
  \frac{iW(y)\Phi'(y)}{E+\Phi(y)}-\frac{\rho \Phi'(y)}{E+\Phi(y)}))\psi_{2}(y)=0
  \end{split}
\end{equation}
To vanish the coefficient of the first order derivative in (\ref{eqq1}), $W(y)$ should satisfy
\begin{equation}\label{Wy}
  W(y)= i\frac{2F(y)^{2}\rho(y)-F'(y)}{2F(y)^{2}}+i\frac{\Phi'(y)}{2F(y)(E-\Phi(y))},
\end{equation}
then, (\ref{eqq1}) turns into
\begin{equation}\label{eqq3}
  -\psi''_{1}(y)+(F(y)^{2}(\Phi(y)^{2}-E^{2})+\frac{3F'(y)^{2}}{4F(y)^{2}}-\frac{F'(y)\Phi'(y)}{2F(y)(E-\Phi(y))}+\frac{3\Phi'^{2}(y)}{4(E-\Phi(y))^{2}}-
  \frac{F''(y)}{2F(y)}+\frac{\Phi''(y)}{2(E-\Phi(y))})\psi_{1}(y)=0.
\end{equation}
Now we may transform (\ref{eqq3}) into a form of (\ref{c1}) whose solutions are known. Hence we can give an ansatze for each function  $F(y)$ and $\Phi(y)$ as below
\begin{eqnarray}
  F(y) &=& \frac{A \cot\sqrt{\alpha}y}{G(y)} \\
  \Phi(y) &=& E+G(y)
\end{eqnarray}
where $A$ is a constant, (\ref{eqq3}) becomes
\begin{equation}\label{eqq4}
   -\psi''_{1}(y)+\left(A^{2}\cot^{2}\sqrt{\alpha}y-\alpha \csc^{2}\sqrt{\alpha}y+3\alpha \csc^{2}2\sqrt{\alpha}y+2A^{2}E\frac{\cot^{2}\sqrt{\alpha}y}{G(y)}\right)\psi_{1}(y)=0.
\end{equation}
We want to match (\ref{c1}) with (\ref{eqq4}), then,  the solutions  $\psi_{1}(y)$ should be  equal to the solutions of (\ref{c1}) which are $\chi_{1}(r)|_{r=y}$. So,  $G(y)$  is found to be
\begin{equation}\label{G}
  G(y)=\frac{8A^{2}E \cot^{2}\sqrt{\alpha}y}{4(A^{2}-C_2)+(-4A^{2}+4C_1+\alpha)\csc^{2}\sqrt{\alpha}y-3\alpha \sec^{2}\sqrt{\alpha}y},
\end{equation}
where
\begin{eqnarray}
  C_2 &=& c+\frac{\alpha}{4}=\frac{9\alpha}{4} \\
 C_1 &=& \frac{3\alpha}{4}.
\end{eqnarray}
Then, $\Omega(y)$ can be given by
\begin{equation}\label{Omg}
  \Omega(y)=\frac{E+\frac{8A^{2}E \cot^{2}\sqrt{\alpha}y}{4(A^{2}-C_2)+(-4A^{2}+4C_1+\alpha)\csc^{2}\sqrt{\alpha}y-3\alpha \sec^{2}\sqrt{\alpha}y}}{m+\Delta(y)}.
\end{equation}
The spectrum of (\ref{eqq3}) can also be given by (\ref{energy}). Next, we can examine (\ref{eqq2}) because we wonder the form of the function $\Omega(y)$ corresponding to that system. Using $W(y)$ in (\ref{Wy}) for the equation (\ref{eqq2}), we may get
\begin{equation}\label{two}
- \psi''_{2}(y)+\frac{2E\Phi'(y)}{E^{2}-\Phi(y)^{2}}\psi'_{2}(y)+\textbf{V}(y)\psi_{2}(y)=0.
\end{equation}
If the coefficient of the first order term is terminated, then, $\Phi(y)=a=constant$ which leads to $\Omega(y)=1$. Choosing $F(y)=\cot\sqrt{\alpha}y$ gives
\begin{equation}\label{f}
- \psi''_{2}(y) +(E^{2}-a^{2}+(a^{2}-E^{2}-\frac{\alpha}{4})\csc^{2}\sqrt{\alpha}y+\frac{3\alpha}{4}\sec^{2}\sqrt{\alpha}y)\psi_{2}(y)=0.
\end{equation}
Let us talk about (\ref{f}) more. This system is a different one which is not discussed above and in the previous Sections. On the other hand, there should be some conditions for (\ref{f}) in order to be a physical eigenvalue equation such as $E^{2}-a^{2} <0$. The potential function in (\ref{f}) known as P\"{o}schl-Teller-I \cite{sukhatme} which is introduced as
\begin{equation}\label{pt1}
  V(x)=-(A+B)^{2}+A(A-\beta)\sec^{2}\beta x+B(B-\beta) \csc^{2} \beta x,
\end{equation}
and the solutions of this model is given in \cite{sukhatme} as
\begin{eqnarray}
  E_{n} &=& (A+B+2n\beta)^{2}-(A+B)^{2} \\
  \psi_{n}(z) &=& (1-z)^{\lambda/2}(1+z)^{s/2}P^{(\lambda-\frac{1}{2}, s-\frac{1}{2})}_{n}(z),
\end{eqnarray}
where $z=1-2\sin^{2}\beta x$, $s=A/\beta$, $\lambda=B/\beta$. Then, if $a=\sqrt{2}E+b$ in (\ref{f}) and say
\begin{eqnarray}
  \beta &=& \sqrt{\alpha} \\
  -(A+B)^{2} &=& -b^{2}-2\sqrt{2}bE \\
  A(A-\beta) &=& \frac{3\alpha}{4} \\
  B(B-\beta) &=& (\sqrt{2}E+b)^{2}-E^{2}-\frac{\alpha}{4}
\end{eqnarray}
we obtain
\begin{eqnarray}
  A &=& \frac{3\sqrt{\alpha}}{2} \\
  B &=&-\frac{E^{2}+2\alpha}{4\sqrt{\alpha}} \\
  b &=&  \frac{-4\sqrt{2}E\alpha+\sqrt{E^{4}\alpha+24E^{2}\alpha^{2}+16\alpha^{3}}}{4\alpha}.
\end{eqnarray}
Thus, we know the solutions of (\ref{f}) whose potential parameters are energy dependent now.
%%%%%%%%%%%%%%%%%%%%%%%%%%%%%%%%%%%%%%%%%%%%%%%%%%%%%%%%%%%%%%%%%%%%%%%%%%%%%%%%%%%%%%%%%%%%%%%%%%%%%%%%%%%%%%%%%%%%%%%%%%%%%%%%%%%%%%%%%%%%%%%%%%%%%%%%%
%%%%%%%%%%%%%%%%%%%%%%%%%%%%%%%%%%%%%%%%%%%%%%%%%%%%%%%%%%%%%%%%%%%%%%%%%%%%%%%%%%%%%%%%%%%%%%%%%%%%%%%%%%%%%%%%%%%%%%%%%%%%%%%%%%%%%%%%%%%%%%%%%%%%%%%%%%
\subsection{ a discussion on  searching the metric function}
For a better understanding, let us discuss a conformally flat space metric $ds^{2}=(1+\mu q^{2})dq^{2}$ whose function  appears as a mass function in the kinetic energy operator $\frac{1}{2}(1+\mu q^{2})^{-1} \textbf{p}^{2}$ \cite{bal}. Because the coefficient of the momentum operator (\ref{XP2}) takes our attention to the position mass dependent Hamiltonians \cite{midya}. If we devote our interest in (\ref{1}) again, this system can be converted into the eigenvalue equation which is Equation $(21)$ in \cite{midya} found by Cari\~{n}ena et al \cite{carinena} previously. In this work, the authors study
\begin{equation}\label{m1}
  -(1+\Lambda y^{2})\psi''(y)-\Lambda y \psi'(y)+(1+\Lambda)\frac{y^{2}}{1+\Lambda y^{2}}\psi(y)=2\varepsilon \psi(y)
\end{equation}
whose solutions are found to be
\begin{equation}\label{df}
  \psi_{m}(y,\Lambda)= H_{m}(y,\Lambda)(1+\Lambda y^{2})^{-\frac{1}{2\Lambda}},~~~~\Lambda > 0
\end{equation}
\begin{equation}\label{ep}
  \varepsilon_{m}=(m+\frac{1}{2})-\frac{m^{2}}{2}\Lambda,~~~~m=0,1,2,...N_{A}
\end{equation}
where $N_{A}$ stands for the greatest integer lower then $m=\frac{1}{\Lambda}$ and $H_{m}(z)$ are the $\Lambda$- deformed Hermite polynomials. Matching (\ref{m1}) with (\ref{1}) gives us the potential functions $U_1(x)$ and $U_2(x)$ which are energy dependent
\begin{eqnarray}
  U_1(x) &=& \frac{E^{2}+C_1 E (1+\alpha x^{2})^{1/2+\gamma/\alpha}+\gamma+(1+\alpha-\gamma^{2})x^{2}}{E(-1+E-C_1(1+\alpha x^{2})^{1/2+\gamma/\alpha})}  \\
  U_2(x) &=& C_1 (1+\alpha x^{2})^{\frac{\alpha+2\gamma}{2\alpha}}-E (1+\alpha x^{2})^{-\frac{1}{2}-\frac{\gamma}{\alpha}+\frac{\alpha+2\gamma}{2\alpha}}.
\end{eqnarray}
Then, solutions of (\ref{1}) can be written as
\begin{eqnarray}\label{En}
  E_{m} &=& \pm \sqrt{\frac{m+\frac{1}{2}-\frac{\alpha m^{2}}{2}}{2}},~~~~m=0,1,2,... \\
  \phi_{1,m}(x,\alpha) &=& H_{m}(x,\alpha)(1+\alpha x^{2})^{-\frac{1}{2\alpha}}.\label{fi}
\end{eqnarray}
In the literature, (\ref{m1}) are known as quantum nonlinear oscillator \cite{Axel}, \cite{Hall}, \cite{Carinena}.

\begin{figure}[h]
\begin{center}
\epsfig{file=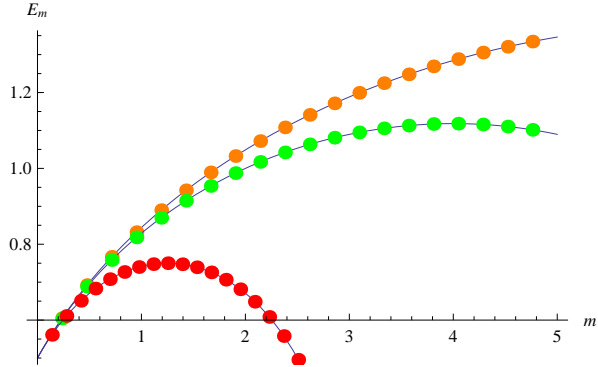,width=7.8cm}
\caption{The energy eigenvalues (\ref{En}) for the values of  $\alpha=0.15$ (orange), $\alpha=0.25$(green), $\alpha=0.8$(red).}
\label{fig1}
\end{center}
\end{figure}
\begin{figure}[h]
\begin{center}
\epsfig{file=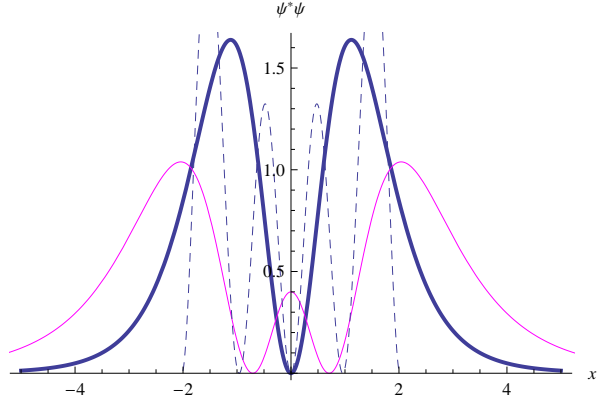,width=7.8cm}
\caption{The probability density for  (\ref{fi}) for the values of  $\alpha=0.2$ and $m=1$ (thick), $m=2$(magenta), $m=5$(dashed).}
\label{fig1}
\end{center}
\end{figure}
For the increasing values of the minimal length parameter $\alpha$, energy eigenvalues decrease and for the greatest $\alpha=0.8$ here, there are only three eigenvalues for $m=0$, $m=1$ and $m=2$. Using the solutions (\ref{fi}), one can obtain the probability densities for a specific value of $\alpha$.

Consider $H_{1}$ given in (\ref{H1}). We have also functions given by (\ref{U1}) and (\ref{yutu}). Let us think about a massless field $m=0$. In that case, vector and scalar potentials become
\begin{eqnarray}
  V(r) &=& \frac{1+E}{2}-\frac{a}{2}\sin\sqrt{\alpha}r-\frac{c+E}{2E(1-E+a\sin\sqrt{\alpha}r)} \\
  S(r) &=& \frac{1-E}{2}+\frac{a}{2}\sin\sqrt{\alpha}r-\frac{c+E}{2E(1-E+a\sin\sqrt{\alpha}r)}.
\end{eqnarray}
In \cite{carlos}, $(1+1)$ dimensional Dirac equation for a free particle in the background of a curved spacetime is given as
\begin{equation}\label{c11}
  i(\frac{\partial}{\partial t}+\frac{\dot{\Omega}}{2\Omega}) \mathrm{\psi} =-i\sigma_{1}(\frac{\partial}{\partial r}+\frac{\Omega'}{\Omega}) \psi
\end{equation}
where $\sigma_{1}$ is the Pauli matrice, $'$ denotes the partial derivative with respect to position $r$ and $\Omega$ is the function of the metric which is  (\ref{ds}). For the static spacetime case, taking $\psi(r,t)=e^{-iEt}[\psi_1~~~~\psi_2]^{T}$, (\ref{c11}) turns into \cite{carlos}
\begin{equation}\label{c2}
  H\psi=\left(
          \begin{array}{cc}
            0 & -i\frac{d}{dr}+V(r) \\
            -i\frac{d}{dr}+V(r) & 0 \\
          \end{array}
        \right)\left(
                 \begin{array}{c}
                   \psi_1(r) \\
                   i\psi_2(r) \\
                 \end{array}
               \right)=E\left(
                          \begin{array}{c}
                            \psi_1(r) \\
                            i\psi_2(r) \\
                          \end{array}
                        \right)
\end{equation}
where \cite{carlos}
\begin{equation}\label{c3}
  \mathcal{V}(r)=-i\frac{\Omega'(r)}{2\Omega(r)}.
\end{equation}
A similarity transformation which is given below
\begin{equation}\label{st}
 \bar{H}= U H U^{-1},~~~~U=\left(
                             \begin{array}{cc}
                               e^{s(r)} & 0 \\
                               0 & e^{s(r)} \\
                             \end{array}
                           \right)
\end{equation}
leads to
\begin{eqnarray}
  \psi'_{2}(r)+(iV(r)-s'(r))\psi_{2}(r) &=& E\psi_{1}(r) \\
  -\psi'_{1}(r)+(-iV(r)+s'(r))\psi_{1}(r) &=& E\psi_{2}(r).
\end{eqnarray}
We can get one of the differential equations
\begin{equation}\label{sson}
  -\psi''_{1}(r)+2(s'(r)-2iV(r))\psi'_{1}(r)+(V(r)^{2}-iV'(r)+2iV(r)s'(r)-s'(r)^{2}+s''(r))\psi_{1}(r)=E^{2}\psi_{1}(r).
\end{equation}
Now, let us match (\ref{sson}) with (\ref{eqq3}), and find
\begin{eqnarray}\label{sonV}
  V(r) &=& -\frac{3i\sqrt{\alpha}}{2} \cot\sqrt{\alpha}r\\
  s(r) &=& \frac{3}{4}\log(-1+\cos2\sqrt{\alpha}r). \label{sonS}
\end{eqnarray}
Thus, for the value of (\ref{sonS}), we can get the function $\Omega(r)$ in (\ref{c11})
\begin{equation}\label{omgson}
  \Omega(r)=k (\sin\sqrt{\alpha}r)^{3},
\end{equation}
$k$ is a constant. Thus, we have extended the solutions given in \cite{carlos}. As it is mentioned in \cite{carlos}, the local phase transformations in spinor solutions may lead to a particular spacetime. On the other hand, for the different position dependent metrics, different effective potentials can be generated \cite{filho}. For the one dimensional motion of the particle, the commutation relation, Weyl-Heisenberg algebra can be given by  \cite{filho}
\begin{equation}\label{f}
  [x,P_{x}]=i\hbar g^{-1/2}_{xx}(x),
\end{equation}
and (\ref{XP}) leads to
\begin{equation}\label{mtr}
  g_{xx}=\frac{1}{(1+\alpha x^{2})^{2}}.
\end{equation}
Here,  the metric  leads to a minimum value for the position uncertainty $\Delta x$, it is known that we  cannot localise a   certain amount of energy in a region smaller than the one defined by its gravitational radius. If  $\Delta p$ is chosen to be very large, then the gravitational field of the particle  makes the   position very imprecise.  Remembering $\Delta x \approx L_{p}$ (Planck length) and $\Delta p \approx \frac{E_{p}}{c}$ (Planck energy/c),  one can give $ds^{2}=\frac{E^{2}_{p}}{c^{2}}dt^{2}-dx^{2}$. Deformed commutation relation also results a force exerted on the particle because of the metric. Then, this makes the particles experience different effective potentials because of the applied force.

\newpage
\section{Conclusions}
We have constructed the connection between the Dirac Hamiltonians in the presence of minimal length and gravitational effects which devotes us to get the solutions of both systems. Moreover, Dirac Hamiltonians are factorized in order to obtain partner potentials of our systems, on the top of that, the transformations between the Dirac systems in flat and curved spacetime may also give a shape function of the metric. This shows that one can use point tarnsformations to obtain a shape function which is non-constant and energy dependent as given in (\ref{Omg}). Considering the nonlinear oscillators mentioned in Section $3.2$, we have extended this study to get different metric solutions for the different potential models. Finally we have obtained another shape function for the system studied in \cite{carlos} which is a special case of our Dirac systems.

\section{Acknowledge}
The author wishes to thank Dr. Altu\~{g} Arda for bringing the topic to her attention.

\end{sloppypar}

\end{document}